\documentclass[conference, 10pt]{IEEEtran}
\IEEEoverridecommandlockouts 
\usepackage[]{graphicx}
\usepackage{caption}
\usepackage{subfig} 
\usepackage{amsmath}
\usepackage{amssymb}
\usepackage{epsfig}
\usepackage{cite}
\usepackage{color}
\usepackage{balance}
\usepackage{amsmath}
\usepackage{amsfonts} 
\usepackage{graphicx}
\usepackage{pdfpages}
\usepackage[T1]{fontenc} 
\usepackage{array}
\usepackage{url}

\usepackage{color}
\usepackage{wrapfig}
\usepackage{lipsum}
\usepackage[hidelinks]{hyperref}
\usepackage{multirow}
\usepackage{tabularx}
\usepackage{tikz}
\usepackage{nth} 
\usepackage{algpseudocode} 
\usepackage{algorithm,algpseudocode}
\usepackage{breqn}
\usepackage{todonotes}
\usepackage{verbatim}

\begin{document}
\title{Monitoring Atmospheric Pollutants From Ground-based Observations}
\author{\IEEEauthorblockN{
Nicholas~Danesi\IEEEauthorrefmark{1},
Mayank~Jain\IEEEauthorrefmark{2}\IEEEauthorrefmark{3},
Yee Hui Lee\IEEEauthorrefmark{4}, and
Soumyabrata Dev\IEEEauthorrefmark{2}\IEEEauthorrefmark{3}
}
\IEEEauthorblockA{\IEEEauthorrefmark{1} School of Biology and Environmental Science, University College Dublin, Ireland}
\IEEEauthorblockA{\IEEEauthorrefmark{2} ADAPT SFI Research Centre, Dublin, Ireland}
\IEEEauthorblockA{\IEEEauthorrefmark{3} School of Computer Science, University College Dublin, Ireland}
\IEEEauthorblockA{\IEEEauthorrefmark{4} School of Electrical and Electronic Engineering, Nanyang Technological University (NTU), Singapore}
\thanks{The ADAPT Centre for Digital Content Technology is funded under the SFI Research Centres Programme (Grant 13/RC/2106\_P2) and is co-funded under the European Regional Development Fund.}
\thanks{Send correspondence to S.\ Dev, E-mail: soumyabrata.dev@ucd.ie.}
\vspace{-0.6cm}
}

\maketitle

\begin{abstract}
Remote sensing analysts continuously monitor the amount of pollutants in the atmosphere. They are usually performed via satellite images. However, these images suffer from low temporal and low spatial resolution. Therefore, observations recorded from the ground offer us a fantastic alternative. There are low-cost sensors that continuously record the PM$_{2.5}$ and PM$_{10}$ concentration levels in the atmosphere. In this position paper, we provide an overview of the state-of-the-art techniques for pollutant forecasting. We establish the interdependence of meteorological parameters on atmospheric pollutants. Our case study in this paper is based on the island of Australia. 
\end{abstract}

\IEEEpeerreviewmaketitle

\section{Introduction}
The increasingly fast-paced development, urbanization, and industrialization of nations around the world has put significant strain on the environment. 
As a result, air quality has significantly deteriorated. 
Air quality degradation is a serious issue especially in heavily urbanized and industrialized areas with significant pollutant emissions~\cite{kaloni2021impact}. World Health Organisation data indicates that air quality in most cities fails to meet safe guidelines. Major air pollutants such as SO$_2$, O$_3$ and particulate matter have serious long-term exposure effects such as heart disease, nerve damage, lung cancer, and respiratory diseases such as emphysema. The short-term effects are also serious, and can include irritation to the nose, throat, eyes, or skin and illnesses such as pneumonia or bronchitis. Even limited exposure to air pollutants is a significant detriment to health. Therefore, there is a clear need to provide accurate short-term ($1$-$4$ hours) forecasting of atmospheric air pollutants for early warning purposes. This will prevent citizens being exposed to dangerous air quality conditions, and 
provide citizens with warnings and health recommendations. 

In this paper, we discuss the effectiveness of data-driven learning models to predict air pollutant concentrations on a short-term basis using historical Australian ground- based meteorological and air pollutant data. This position paper also aims to identify the key meteorological variables that affect atmospheric air pollutant concentrations and examine the forecasting model's effectiveness in a variety of climates including temperate, tropical, subtropical, and arid. The results from this position paper will help facilitate authorities to provide accurate early warnings for air quality to citizens and to help industrial emitters adjust their production to maintain pollutant emission compliance to better regulate air quality. Our results will also fill a research gap, as there has been no prior research in Oceania on utilizing deep learning models for forecasting air quality.

\vspace{-0.1cm}

\section{Monitoring Atmospheric Pollutants}

Air quality forecasting and early warning systems are essential for health authorities to provide citizens with warnings about low air quality and provide health recommendations. Historically, there has not been a reliable forecasting model, which has led to the development of several types of forecasting methods.

\subsection{Data-driven techniques}
The traditional methods for forecasting include statistical forecasting, artificial intelligence methods and numerical models. Most models are statistical and are limited in range and effectiveness. In \cite{bai2018air}, Bai \textit{et al.} have provided a overview of the different air pollution forecasting methods. The authors have discussed the various data-driven prediction techniques including regression methods, autoregressive integrated moving average (ARIMA) models, Projection Pursuit Model (PP) models, and Principal Component Analysis (PCA) models~\cite{manandhar2019data}. These methods require low computational time in predicting future values. However, they have poor performance as compared to deep neural networks.

\subsection{Deep-learning models}
In recent years, Long-Short-Term Memory (LSTM) network models have been used considerably for effective times series forecasting. 
LTSM deep learning models for short-term air quality forecasting have been successfully tested in several case studies in China~\cite{liu2020air}, India~\cite{rao2019air}, and South Korea~\cite{kim2019development}. Liu \textit{et al.} proposed a hybrid prediction model using wavelet decomposition, LSTM network and information gain to predict atmospheric pollutants in Beijing~\cite{liu2020air}. Rao \textit{et al.} in \cite{rao2019air} also used LSTM networks effectively to predict the air quality for the city of Visakhapatnam in India. The researchers obtained good prediction accuracy in the hourly based air ambiance values. Similarly, a deep neural network based on LSTM network was developed by Kim \textit{et al.} for daily predictions of PM$_{10}$ and PM$_{2.5}$ in South Korea~\cite{kim2019development}.

\subsection{Case study in Australia}
Australia is a country with high density urban cities as well as areas with heavy industrial emitters of air pollutants~\cite{dean2017climate}, so by providing authorities with accurate short-term forecasting of air pollutants, early health warnings can be provided to citizens. Early warning systems also allows industrial emitters to adjust production to accommodate changing air quality conditions. As of the 29th of October 2020, Australian state meteorological departments do not use LTSM air quality forecasting as confirmed by phone through the Commonwealth Scientific and Industrial Research Organisation (CSIRO). Because air pollution levels are highly correlated with climatic and meteorological conditions, Australia's distinct climates provide an exceptionally good opportunity for us as a case study. 

\section{Results \& Discussion}

\subsection{Dataset}
In this paper, we obtain time-stamped data from several ground monitoring stations that are located across Australia. We systematically scrap the pollutant data from the Queensland Government website\footnote{The meteorological data along with the pollutant concentrations is obtained from \url{https://apps.des.qld.gov.au/air-quality/download/}.}. The collected data contains the daily average of the various meteorological parameters and the pollutants. The meteorological variables include air temperature, wind velocity and direction, relative humidity, air pressure, and rainfall. The pollutant concentrations comprises the daily averages of PM$_{2.5}$, PM$_{10}$, O$_3$, NO$_2$, and SO$_2$. We obtain these observations for the yearly duration of the year $2019$.

\subsection{Interdependence of meteorological parameters}
\begin{figure}[htb]
\centering
\includegraphics[width=0.50\textwidth]{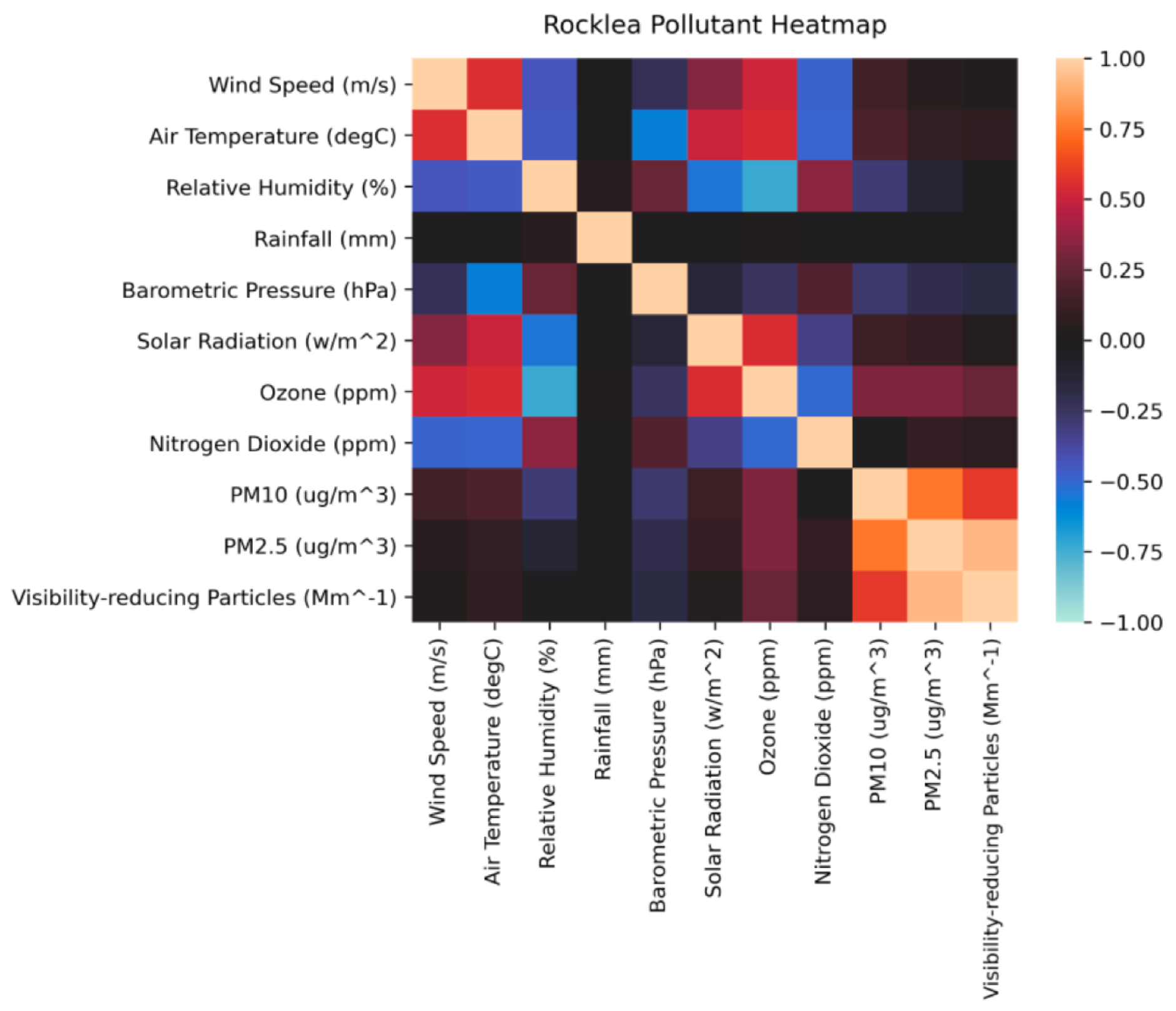} 
\caption{We demonstrate the relationship of the various meteorological parameters on the pollutant concentrations.}
\label{fig:heatmap}
\vspace{-0.5cm}
\end{figure}

In this paper, we focus our analysis on the Rocklea weather station, based in the suburbs of Queensland, Australia. Figure~\ref{fig:heatmap} demonstrates the relationship amongst the meteorological parameters and pollutant concentrations. We observe that the pollutant O$_3$ is strongly correlated with solar radiation and air temperature. Similarly, the other pollutants PM$_{10}$, PM$_{2.5}$ and visibility reducing particles are strongly correlated amongst each other. These pollutants have low correlation with air temperature, wind speed and relative humidity. Furthermore, Fig.~\ref{fig:heatmap} also describes the inter-relationship amongst the meteorological parameters. We observe that solar radiation and air temperature are strongly positively correlated. Similarly, air temperature and barometric pressure are strongly negatively correlated, and so is solar radiation and relative humidity that is strongly negatively correlated. These observations are similar for the other weather stations located at other locations of Australia.

\section{Conclusion \& Future Work}
In this paper, we described an approach for continuous monitoring of the atmospheric pollutants using low-cost ground-based observation stations. Our proposition is that we leverage on best practices of systematic data analysis and deep learning techniques on collected data source to understand pollutant concentrations in the atmosphere. This paves path for us to devise deep neural network for accurate short-term forecasting of the pollutant concentrations. This will act as a starting point for crystallizing our vision for a reliable and accurate pollutant forecasting, enabling for early warning systems for the citizens. Our case study is based in Oceania. Our future works of this position paper include a detailed analysis of pollutant concentration across varying climate conditions and larger statistical duration. Using LSTM networks~\cite{jain2020forecasting}, we also plan to propose a data-driven framework for reliable pollutant forecasting.

\vspace{-0.3cm}


\bibliographystyle{IEEEtran.bst}

\begin{thebibliography}{1}
\providecommand{\url}[1]{#1}
\csname url@samestyle\endcsname
\providecommand{\newblock}{\relax}
\providecommand{\bibinfo}[2]{#2}
\providecommand{\BIBentrySTDinterwordspacing}{\spaceskip=0pt\relax}
\providecommand{\BIBentryALTinterwordstretchfactor}{4}
\providecommand{\BIBentryALTinterwordspacing}{\spaceskip=\fontdimen2\font plus
\BIBentryALTinterwordstretchfactor\fontdimen3\font minus
  \fontdimen4\font\relax}
\providecommand{\BIBforeignlanguage}[2]{{%
\expandafter\ifx\csname l@#1\endcsname\relax
\typeout{** WARNING: IEEEtran.bst: No hyphenation pattern has been}%
\typeout{** loaded for the language `#1'. Using the pattern for}%
\typeout{** the default language instead.}%
\else
\language=\csname l@#1\endcsname
\fi
#2}}
\providecommand{\BIBdecl}{\relax}
\BIBdecl

\bibitem{kaloni2021impact}
D.~Kaloni, Y.~H. Lee, and S.~Dev, ``Impact of {COVID19}-induced lockdown on air
  quality in {Ireland},'' in \emph{Proc. International Geoscience and Remote
  Sensing Symposium (IGARSS)}, 2021.

\bibitem{bai2018air}
L.~Bai, J.~Wang, X.~Ma, and H.~Lu, ``Air pollution forecasts: An overview,''
  \emph{International journal of environmental research and public health},
  vol.~15, no.~4, p. 780, 2018.

\bibitem{manandhar2019data}
S.~Manandhar, S.~Dev, Y.~H. Lee, Y.~S. Meng, and S.~Winkler, ``A data-driven
  approach for accurate rainfall prediction,'' \emph{IEEE Transactions on
  Geoscience and Remote Sensing}, vol.~57, no.~11, pp. 9323--9331, 2019.

\bibitem{liu2020air}
B.~Liu, X.~Guo, M.~Lai, and Q.~Wang, ``Air pollutant concentration forecasting
  using long short-term memory based on wavelet transform and information gain:
  A case study of {Beijing},'' \emph{Computational Intelligence and
  Neuroscience}, vol. 2020, 2020.

\bibitem{rao2019air}
K.~S. Rao, G.~L. Devi, and N.~Ramesh, ``Air quality prediction in
  {Visakhapatnam} with {LSTM} based recurrent neural networks,''
  \emph{International Journal of Intelligent Systems and Applications},
  vol.~11, no.~02, p. 2019, 2019.

\bibitem{kim2019development}
H.~S. Kim, I.~Park, C.~H. Song, K.~Lee, J.~W. Yun, H.~K. Kim, M.~Jeon, J.~Lee,
  and K.~M. Han, ``Development of a daily {PM}$_{10}$ and {PM}$_{2.5}$
  prediction system using a deep long short-term memory neural network model,''
  \emph{Atmospheric Chemistry and Physics}, vol.~19, no.~20, pp.
  12\,935--12\,951, 2019.

\bibitem{dean2017climate}
A.~Dean and D.~Green, ``Climate change, air pollution and health in
  {Australia},'' \emph{Sydney, Australia: {UNSW} Sydney}, 2017.

\bibitem{jain2020forecasting}
M.~Jain, S.~Manandhar, Y.~H. Lee, S.~Winkler, and S.~Dev, ``Forecasting
  precipitable water vapor using {LSTM}s,'' in \emph{Proc. IEEE AP-S Symposium
  and USNC-URSI Radio Science Meeting}, 2020.

\end{thebibliography}

\end{document}